\documentstyle[epsfig,12pt]{article}
\def\lsim{\mathrel{\rlap{\lower4pt\hbox{\hskip1pt$\sim$}}
    \raise1pt\hbox{$<$}}}               
\def\gsim{\mathrel{\rlap{\lower4pt\hbox{\hskip1pt$\sim$}}
    \raise1pt\hbox{$>$}}}               

\newcommand{\be}{\begin{eqnarray}}
\newcommand{\ee}{\end{eqnarray}}

\begin{document}

\rightline{\Large{Preprint RM3-TH/99-10}}

\vspace{2cm}

\begin{center}

\Large{Analysis of the exclusive semileptonic decay $\Lambda_b \to 
\Lambda_c + \ell \bar{\nu}_{\ell}$\\[3mm] within a light-front constituent 
quark model\footnote{\bf In the Proceedings of the International Conference 
PANIC '99, Uppsala (Sweden), June 1999, to appear in Nuclear Physics A.}}

\vspace{2cm}

\large{F. Cardarelli and S. Simula}

\vspace{1cm}

\normalsize{Istituto Nazionale di Fisica Nucleare, Sezione Roma III\\
Via della Vasca Navale 84, I-00146, Roma, Italy}

\vspace{1cm}

\begin{abstract}

\indent The exclusive semileptonic decay $\Lambda_b \to \Lambda_c + \ell 
\bar{\nu}_{\ell}$ is investigated including both radiative and first-order 
power corrections in the inverse heavy-quark mass, while the Isgur-Wise 
function is calculated within a light-front constituent quark model. It 
turns out that the dependence on the model parameters can be effectively 
constrained using recent lattice $QCD$ results at low values of the recoil. 
Our final predictions for the exclusive semileptonic branching ratio, the 
longitudinal and transverse asymmetries, and the longitudinal to transverse 
decay ratio are: $Br(\Lambda_b \to \Lambda_c \ell \bar{\nu}_{\ell}) = (6.3 
\pm 1.6) ~ \% ~ |V_{bc} / 0.040|^2 ~ \tau(\Lambda_b) / (1.24 ~ ps)$, $a_L = 
-0.945 \pm 0.014$, $a_T = -0.62 \pm 0.09$ and $R_{L/T} = 1.57 \pm 0.15$, 
respectively. Moreover, the theoretical uncertainties both on $a_L$ and the 
(partially integrated) $R_{L/T}$ are found to be quite small and, therefore, 
the experimental determination of these quantities is a very interesting tool 
for testing the Standard Model and for investigating possible New Physics. In 
this respect the sensitivity to extract unique information both on the 
strength and phase of possible hadronic right-handed currents is illustrated.

\end{abstract}

\end{center}

\newpage

\rightline{}

\newpage

\setcounter{page}{1}

\pagestyle{plain}

\section{INTRODUCTION}

\indent The investigation of the exclusive semileptonic decays of heavy
 hadrons, driven by the elementary process $b \to c + \ell \bar{\nu}_{\ell}$, 
can provide relevant information on fundamental parameters of the Standard 
Model ($SM$), like e.g. the quark masses and the weak mixing angle $V_bc$, 
as well as on possible extensions of the $SM$. In this contribution we 
present the main results of an analysis of the decay $\Lambda_b \to 
\Lambda_c + \ell \bar{\nu}_{\ell}$ performed recently in Refs. 
\cite{CAR98,CAR99} adopting a relativistic quark model formulated on 
the light-front, and we illustrate also the possibility to extract 
relevant information on hadronic right-handed currents from the $\Lambda_b 
\to \Lambda_c + \ell \bar{\nu}_{\ell}$ process.

\indent In Refs. \cite{CAR98,CAR99} the Isgur-Wise ($IW$) form factor 
relevant for the $\Lambda_b \to \Lambda_c + \ell \bar{\nu}_{\ell}$ decay  
has been calculated in the whole accessible kinematical range adopting a 
light-front constituent quark model and using various forms of the 
three-quark wave function. It turns out that the $IW$ form factor is 
sensitive to relativistic delocalization effects associated to the 
light-quark degrees of freedom, leading to a saturation property of 
the $IW$ form factor as a function of the canonical baryon size. 
Moreover, the shape of the $IW$ function is found to be significantly 
affected by the baryon structure, being sharply different in case of 
diquark-like or collinear-type configurations. The comparison with recent 
lattice $QCD$ calculations \cite{UKQCD} as well as with the 
(model-dependent) dispersive bounds of Ref. \cite{gupta} suggests 
clearly the dominance of collinear-type configurations with respect 
to diquark-like ones, and allows to put effective {\em constraints} on 
the shape of the $IW$ function in the full recoil range relevant for 
the $\Lambda_b \to \Lambda_c + \ell \bar{\nu}_{\ell}$ decay (see Ref. 
\cite{CAR99} for more details).

\section{ANALYSIS OF THE $\Lambda_b \to \Lambda_c + \ell \bar{\nu}_{\ell}$ 
DECAY WITHIN THE SM}

\indent The $\Lambda_b \to \Lambda_c + \ell \bar{\nu}_{\ell}$ decay has 
been investigated within the framework of the $SM$ including radiative 
as well as first-order $1 / m_Q$ corrections to the relevant form factors, 
as derived in Ref. \cite{neubert}. Our final predictions for the exclusive 
semileptonic branching ratio, the longitudinal and transverse asymmetries, 
and the longitudinal to transverse decay ratio are: $Br_{SL} = (6.3 \pm 1.4) 
~ \% ~ |V_{bc} / 0.040|^2 ~ \tau(\Lambda_b) / (1.24 ~ ps)$, $a_L = -0.94 
\pm 0.01$, $a_T = -0.62 \pm 0.08$ and $R_{L/T} = 1.57 \pm 0.12$, 
respectively. Our results for the partially-integrated exclusive 
semileptonic branching ratio are reported in Fig. 1 and compared with 
the corresponding results from the lattice $QCD$ simulations of Ref. 
\cite{UKQCD}. It can be seen that our results are always well within 
the range of values given by the lattice $QCD$ simulations and that 
radiative plus first-order power corrections modify only slightly the 
results obtained within the Heavy Quark Symmetry ($HQS$). If the 
integration over the recoil is limited to $\omega = 1.2$, then the 
resulting uncertainty on the branching ratio reduces significantly 
to $\simeq 12 \%$, though at the price of reducing the number of the 
events by a factor $\simeq 0.44$ (see Table 2 of Ref. \cite{CAR99}). 
This implies the possibility to extract the $CKM$ matrix elements 
$|V_{bc}|$ with a theoretical uncertainty of $\simeq 6 \%$, which 
is comparable with present uncertainties obtained from exclusive 
semileptonic $B$-meson decays \cite{bigi}.

\begin{figure}[htb]

\parbox{9.25cm}{\epsfxsize=9cm \epsfig{file=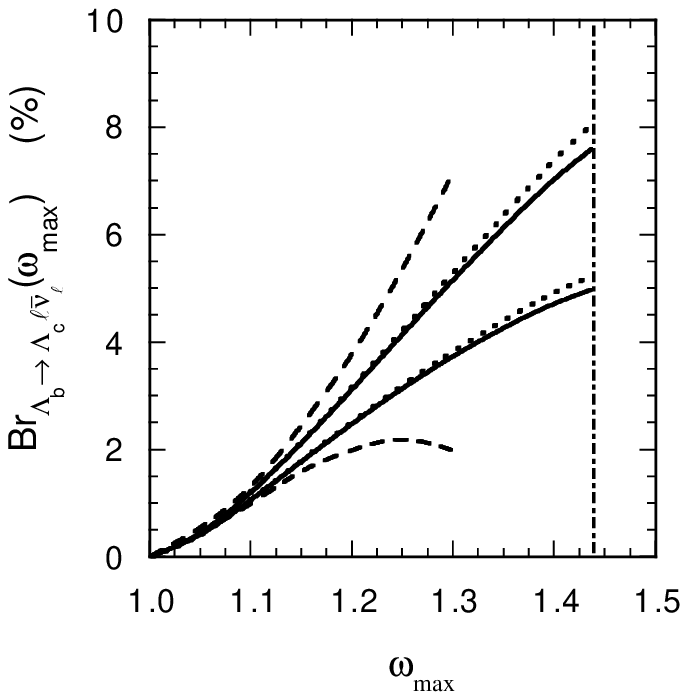}} \ $~~$ \ 
\parbox{6.5cm}{\small \noindent Figure 1. Partially-integrated branching 
ratio $Br_{\Lambda_b \to \Lambda_c \ell \bar{\nu}_{\ell}}(\omega_{max})$ 
in $\%$ versus the upper limit of integration over the recoil, 
$\omega_{max}$, calculated at $|V_{bc}| = 0.040$ and $\tau(\Lambda_b) = 
1.24 ~ ps$ \cite{PDG98}. The solid and dashed lines correspond to our 
and lattice $QCD$ results \cite{UKQCD}, respectively. The lower and 
upper solid lines are the results corresponding to the lower and upper 
limits of the $IW$ function derived in Ref. \cite{CAR99}, including 
radiative plus first-order $1 / m_Q$ corrections. The dotted lines 
are the $HQS$ results. The vertical dot-dashed line indicates the 
physical threshold $\omega_{th} \simeq 1.44$. (After Ref. \cite{CAR99}).}

\vspace{0.5cm}

\end{figure}

\indent In Ref. \cite{CAR99} various partially-integrated asymmetries 
have been investigated, showing that in comparing with (future) data 
the precise $\omega$-range of the experiments has to be taken into account. 
It turns out that radiative plus first-order $1 / m_Q$ corrections are 
relevant for the transverse asymmetry $a_T$ and for the longitudinal 
$\Lambda_c$ polarization $P_L$, whereas both $a_L$ and $R_{L/T}$ are 
only marginally affected by such $HQS$ corrections, as it is clearly 
shown in Fig. 2. Moreover, the model dependence on the various 
asymmetries is generally quite limited and, in particular, our 
uncertainty on $R_{L/T}$, which is always much less than the one 
presently achievable by lattice $QCD$ calculations, reduces to 
$\simeq 1 \%$ when the integration over the recoil is limited to 
$\omega = 1.2$.

\begin{figure}[htb]

\centerline{\epsfxsize=16cm \epsfig{file=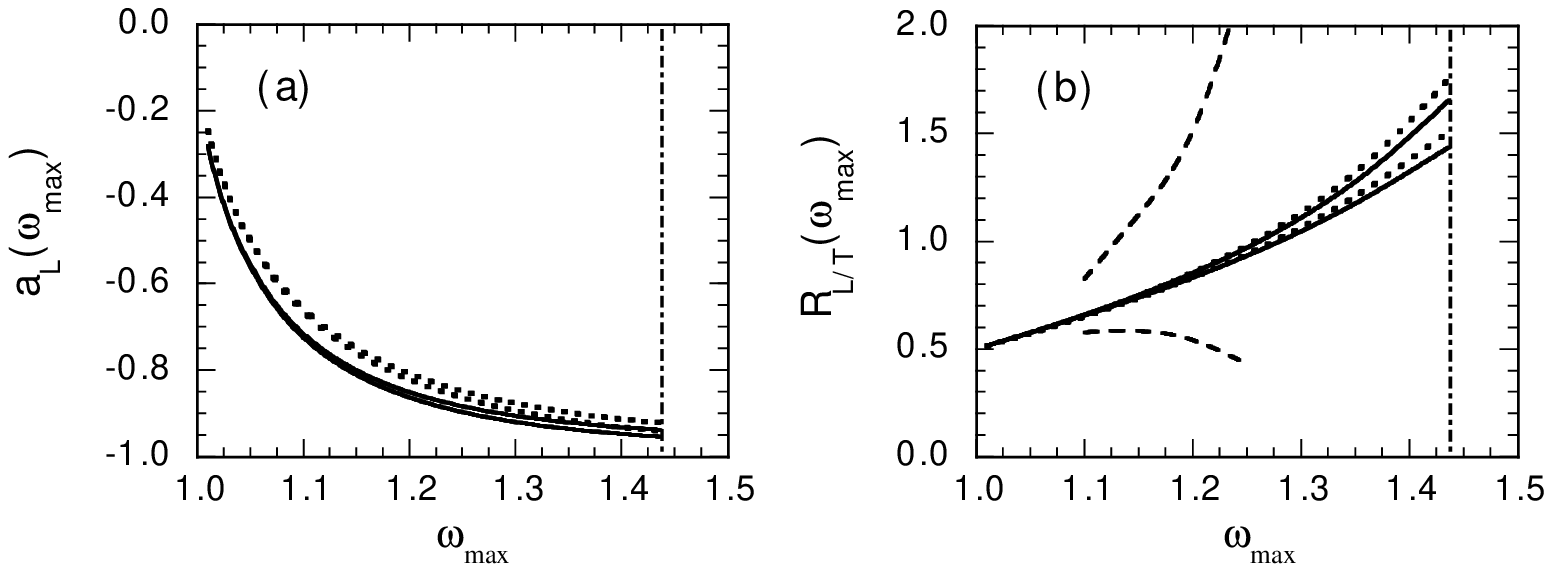}}

{\small \noindent Figure 2. Partially-integrated longitudinal asymmetry 
$a_L$ (a) and longitudinal to transverse decay ratio $R_{L/T}$ (d)
 versus the upper limit of integration over the recoil, $\omega_{max}$, 
for the decay process $\Lambda_b \to \Lambda_c + \ell \bar{\nu}_{\ell}$. 
The dotted and solid lines correspond to the $HQS$ results and to those 
obtained including radiative plus first-order $1 / m_Q$ corrections, 
respectively. The meaning of the lower and upper lines, as well as of 
the vertical dot-dashed line, is as in Fig. 1. In (b) the dashed lines 
correspond to the lattice $QCD$ results of Ref. \cite{UKQCD}. (Adapted 
from Ref. \cite{CAR99}).}

\vspace{0.5cm}

\end{figure}

\section{EFFECTS FROM POSSIBLE RIGHT-HANDED CURRENTS}

\indent The small uncertainties found for the longitudinal asymmetry 
$a_L$ and the longitudinal to transverse decay ratio $R_{L/T}$ make 
the experimental determination of these quantities a very interesting 
tool for testing the $SM$ and investigating possible New Physics ($NP$). 
We now illustrate the sensitivity of $a_L$ and $R_{L/T}$ to the 
introduction of possible hadronic right-handed currents by considering 
the presence in the effective weak Hamiltonian of both left-handed 
($LH$) and right-handed ($RH$) operators, viz. ${\cal{H}}_{eff}(b \to c 
\ell \bar{\nu}_{\ell}) = c_L O_L(b \to c \ell \bar{\nu}_{\ell}) + c_R 
O_R(b \to c \ell \bar{\nu}_{\ell})$, where $O_{L(R)}(b \to c \ell 
\bar{\nu}_{\ell}) \equiv (\bar{c} \gamma_{\mu} (1 \mp \gamma_5) b) 
(\ell \gamma^{\mu} (1 - \gamma_5) \nu_{\ell})$, with the values of 
the coefficients $c_L$ and $c_R$ depending on the specific $NP$ model. 
In general $c_L \neq c_L^{(SM)}$, but in what follows we consider for 
simplicity $c_L = c_L^{(SM)}$, and we introduce two parameters, $\rho$ 
and $\eta$, defined as $\rho \equiv |c_R| / |c_L|$ and $\eta \equiv - 
\mbox{Re}(c_L c_R^*) / (|c_L|^2 + |c_R|^2)$, which are clearly connected 
to the relative strength and phase of $RH$ currents with respect to $LH$ 
ones. On one hand side we have found that the branching ratio $Br_{SL}$ 
is sharply sensitive to the values of $\rho$ and $\eta$; however, the 
$SM$ predictions for $Br_{SL}$ (see Fig. 1) can be obtained also with 
different pairs of values of $(\rho, ~ \eta)$. On the other hand side 
our main results for $a_L$ and $R_{L/T}$ are summarized in Fig. 3. It 
can be clearly seen that both $a_L$ and $R_{L/T}$ are significantly 
sensitive to the presence of $RH$ currents, but their determination 
allows the extraction of information both on the strength and phase 
of $RH$ currents.

\begin{figure}[htb]

\parbox{9.25cm}{\epsfxsize=8cm \epsfig{file=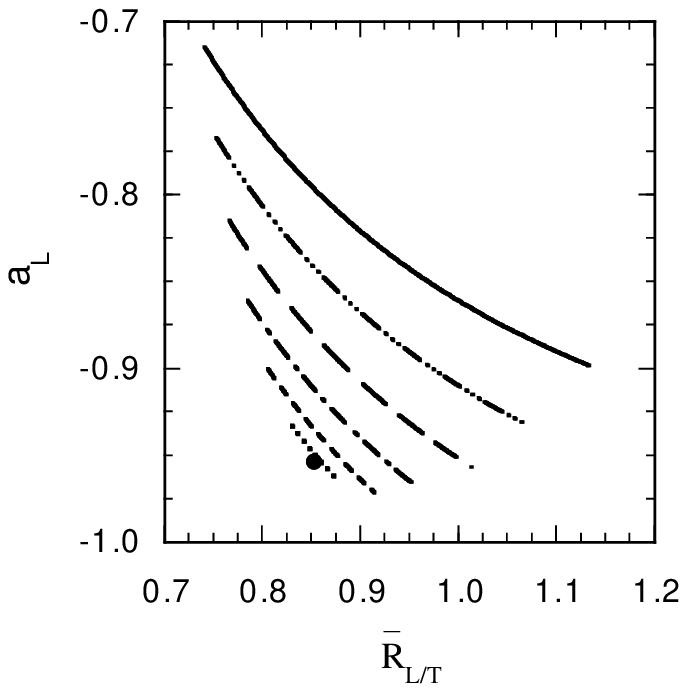}} \ $~~$ \ 
\parbox{6.5cm}{\small \noindent Figure 3. The longitudinal asymmetry $a_L$ 
versus the partially-integrated longitudinal to transverse ratio 
$\bar{R}_{L/T} \equiv R_{L/T}(\omega_{max} = 1.2)$ for the decay process 
$\Lambda_b \to \Lambda_c + \ell \bar{\nu}_{\ell}$. The full dot 
corresponds to the results calculated within the $SM$ framework, while 
the various lines are obtained for fixed values of $\rho$ and varying 
$\eta$ in its allowed range (see text). The dotted, dashed, dot-dashed, 
long-dashed, triple-dotted-dashed and solid lines correspond to $\rho = 
0.05, 0.10, 0.15, 0.20, 0.25$ and $0.30$, respectively.}

\vspace{0.5cm}

\end{figure}

\section{CONCLUSIONS}

\indent The exclusive semileptonic decay $\Lambda_b \to \Lambda_c + \ell 
\bar{\nu}_{\ell}$ has been investigated including both radiative and 
first-order power corrections in the inverse heavy-quark mass, while 
the Isgur-Wise function has been calculated within a light-front 
constituent quark model. The dependence on the model parameters has 
been effectively constrained using recent lattice $QCD$ results at 
low values of the recoil. Our predictions for the exclusive semileptonic 
branching ratio, the longitudinal and transverse asymmetries, and the 
longitudinal to transverse decay ratio are: $Br(\Lambda_b \to \Lambda_c 
\ell \bar{\nu}_{\ell}) = (6.3 \pm 1.6) ~ \% ~ |V_{bc} / 0.040|^2 ~ 
\tau(\Lambda_b) / (1.24 ~ ps)$, $a_L = -0.945 \pm 0.014$, $a_T = -0.62 
\pm 0.09$ and $R_{L/T} = 1.57 \pm 0.15$, respectively. Moreover, the 
theoretical uncertainties both on the longitudinal asymmetry and the 
(partially integrated) longitudinal to transverse decay ratio are found 
to be quite small and, therefore, the experimental determination of 
these quantities is a very interesting tool for testing the Standard 
Model and for investigating possible New Physics. In this respect the 
possibility to extract unique information both on the strength and 
phase of hadronic right-handed currents has been illustrated.

\end{document}